\documentclass[Physsubmission, Phys]{SciPost}

\hypersetup{
    colorlinks,
    linkcolor={red!50!black},
    citecolor={blue!50!black},
    urlcolor={blue!80!black}
}

\usepackage[bitstream-charter]{mathdesign}
\DeclareSymbolFont{usualmathcal}{OMS}{cmsy}{m}{n}
\DeclareSymbolFontAlphabet{\mathcal}{usualmathcal}

\begin{document}

\begin{center}{\Large \textbf{
WG2 Summary: Small-x, Diffraction and Vector Mesons\
}}\end{center}

\begin{center}
Tolga Altinoluk\textsuperscript{1},
Marta Luszczak\textsuperscript{2} and
Daniel Tapia Takaki\textsuperscript{3$\star$}
\end{center}

\begin{center}
{\bf 1} Theoretical Physics Division, National Centre for Nuclear Research,\\
ul. Pasteura 7,  02-093 Warsaw, Poland
\\
{\bf 2} College of Natural Sciences, Institute of Physics, University of Rzeszow,\\
ul. Pigonia 1, PL-35310 Rzeszow, Poland
\\
{\bf 3} Department of Physics and Astronomy, University of Kansas,\\
 Lawrence KS, 66045 USA
\\
* tolga.altinoluk@ncbj.gov.pl, luszczak@ur.edu.pl, Daniel.Tapia.Takaki@cern.ch
\end{center}

\begin{center}
\today
\end{center}


\definecolor{palegray}{gray}{0.95}
\begin{center}
\colorbox{palegray}{
  \begin{tabular}{rr}
  \begin{minipage}{0.1\textwidth}
    \includegraphics[width=22mm]{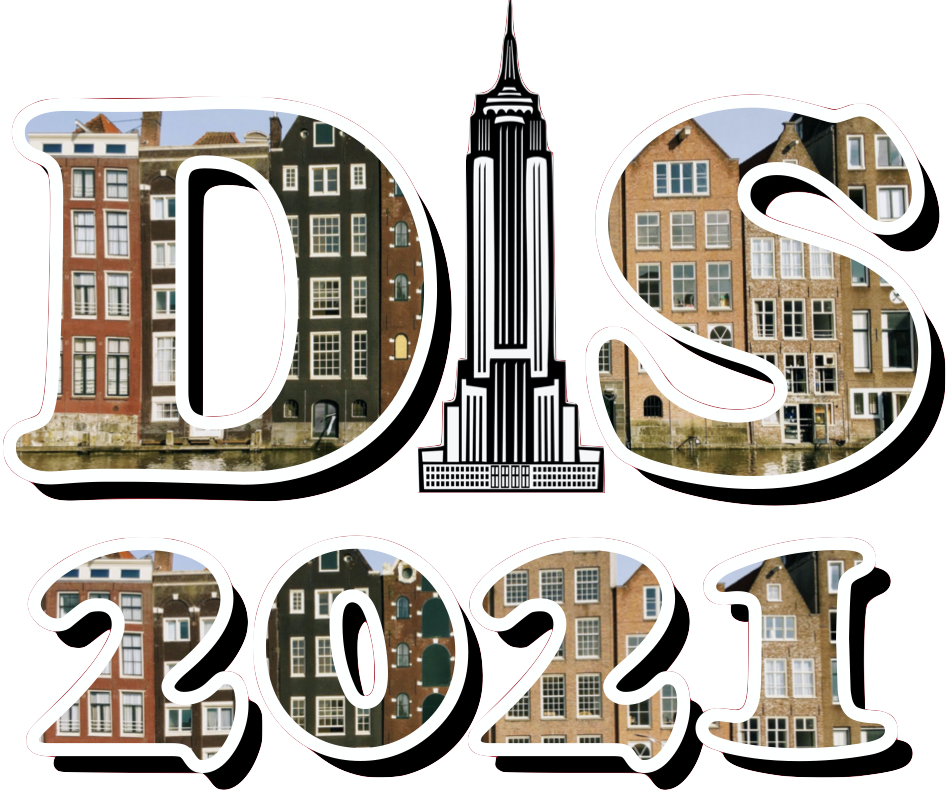}
  \end{minipage}
  &
  \begin{minipage}{0.75\textwidth}
    \begin{center}
    {\it Proceedings for the XXVIII International Workshop\\ on Deep-Inelastic Scattering and
Related Subjects,}\\
    {\it Stony Brook University, New York, USA, 12-16 April 2021} \\
    \doi{10.21468/SciPostPhysProc.?}\\
    \end{center}
  \end{minipage}
\end{tabular}
}
\end{center}

\section*{Abstract}
{\bf
Here we briefly summarize the contributions to Working Group 2: Small-x, Diffraction and Vector Mesons, at the DIS2021 Workshop.
}

\vspace{10pt}
\noindent\rule{\textwidth}{1pt}
\tableofcontents\thispagestyle{fancy}
\noindent\rule{\textwidth}{1pt}
\vspace{10pt}

\section{Introduction}
\label{sec:intro}
The DIS2021 Workshop provided a great opportunity to present and discuss many new ideas and experimental results. In the parallel sessions devoted to Working Group 2: Small-x, Diffraction and Vector Mesons a total of 49 talks (31 theoretical and 18 experimental) were given. The aim of this manuscript is to provide a short summary of the contributions to Working Group 2 and it is organized as follows. In Section \ref{theory_talks}, we summarize the theoretical talks and continue with the summary of the experimental talks in Section \ref{experimental_talks}. Finally, we shortly present conclusions in Section \ref{conclusions}. All talks can be found on the Indico site of the workshop. We refer to the corresponding proceedings contributions of each talk provided that they are available at the time of writing this summary, otherwise we mention the relevant papers for the works that are available online. 

\section{Theory Talks}
\label{theory_talks}
Many exciting results and progress in many theoretical aspects in the field was presented in DIS2021 Workshop. In order to organize the summary of these presentations, theory talks are roughly divided into subject-like subsections which are inclusive production and correlations (subsection \ref{inclusive}), improvements of the perturbative saturation (subsection \ref{improvements}), diffraction (subsection \ref{diffraction}), and exclusive processes and vector meson production (subsection \ref{VM}).

\subsection{Inclusive production and correlations}
\label{inclusive}

One of the most interesting observables that have been studied heavily over the last couple years is the forward dijet production since in the so-called correlation (also known as back-to-back) limit, it allows one to probe the high energy limit of the transverse momentum dependent parton distribution functions (TMDs) from the Color Glass Condensate (CGC) calculations. Moreover, in this limit, one can distinguish between the kinematic power corrections (which leads to so called small-x improved TMD (iTMD) framework when resummed to all orders) and the genuine saturation effects. At DIS2021, we had three talks devoted to the forward multi-jet production. 
F. Salazar, presented his recent work about the gluon saturation effects in forward dijet production at the EIC. He discussed the relative importance of the kinematical power corrections and genuine saturation effects for differential yields and elliptic flow as a function of the total transverse momenta of the produced jets \cite{Boussarie:2021lkb}. P. Kotko,  showed a new calculation framework that combines the iTMD framework and Sudakov resummation and discussed the angular correlations between the dijet pair and outgoing electron as well as the angular correlations between the produced dijets within this new framework for DIS at EIC \cite{vanHameren:2021sqc}. A. van Hameran, discussed about extension of the iTMD formulation (which was originally derived for dijet production) to the case of trijet production and presented his results both for proton-proton and for proton-lead collisions at the center of mass energy $5.02$ TeV where a significant saturation effects in the nuclear modification factor for momentum imbalance-sensitive observables are shown \cite{Bury:2020ndc}.    

Let us now continue our summary with the talks that were devoted to the correlations at mid rapidity. N. Armesto, presented the results of his recent work on the correlations due the azimuthal asymmetry, specifically the squared second Fourier harmonic $v_2^2$, and the total multiplicity in the event. The correlations were found to be very small which is consistent with the observations.Moreover,  an interesting interplay between the HBT and Bose enhancement effects are noted when the correlations are plotted as a function of the transverse momentum bin width \cite{Altinoluk:2021gsa}. P. Agostini, discussed four gluon production in the dilute-dense CGC framework by using the Wigner function approach to define the color charges in the projectile wave function. His result for the four particle cumulant $c_2\{4\}$ turned out to be negative which provides a sensible second order Fourier coefficient \cite{Agostini:2021xca}. A. Dumitru, discussed the color charge correlations in the proton by using its light-front wave function. The presented results included the impact parameter dependence as well as the mixing between the impact parameter and dipole size, both of which are usually neglected in the standard CGC calculations. The effects of these dependences were compared with the models used in standard CGC \cite{Dumitru:2021hjm}.   

We also had two talks concentrated on the parton distributions at small-x . G. A. Chirilli gave a talk on the small-x behavior of both gluon and quark quasi distributions. He presented the calculation of the small-x behavior of these two parton distributions in two ways, namely in the saddle point approximation and the leading twist approximation. He argued that the correlation function of quark and gluon quasi parton distribution functions agree with the general result for two-point correlation function in conformal field theory. Y. Mehtar-Tani presented a new calculation framework to study the high energy limit of the parton distributions. He showed that a new gauge invariant operator definition of the unintegrated gluon distributions emerges naturally in this framework and it systematically accounts for the collinear limit of the structure functions. He also discussed inclusive DIS as an application of this framework \cite{Boussarie:2020fpb}. 

Finally, we had two talks on the future of small-x physics at the new colliders. A. Stasto, gave a summary of the status of the planned studies for the LHeC and FCC-eh particularly focusing on parton distributions at small-x, potential tests for saturation, longitudinal structure function and diffractive phenomena \cite{LHeC:2020van}. Another future oriented talk was given by J. Ralston. He argued that by using entanglement and quantum tomography, one can go beyond the traditional experimental data analysis procedure of making distributions and cuts. 



\subsection{Improvements of the perturbative saturation framework}
\label{improvements}

The two distinct yet complimentary ways to improve the perturbative saturation framework are to either to go further in the perturbative coupling constant and perform the calculations at next-to-leading order (NLO) in the coupling constant or to stay at leading order but include the corrections beyond the eikonal approximations. We have witnessed a lot of progress in improving the perturbative saturation framework that follows either one of the above mentioned ways over the last years. 

Let us first start with summarizing the talks on NLO studies of different observables. M. Hentschinski discussed NLO corrections to Higgs production in the forward region by using the Lipatov's high energy effective action. He provided a proper definition of the NLO coefficient and also proposed a subtraction mechanism for this coefficient  to have a stable cancellation of real and virtual infra-red singularities \cite{Hentschinski:2020tbi}. On the more phenomenological side, H. Hanninen discussed his results on the first comparison of the NLO DIS cross sections to HERA data. Due to the fact that NLO DIS inclusive cross sections are calculated only in the massless quark limit, he constructed and fitted a dataset for light-quark cross sections using an independent parametrization of HERA total and heavy quark data which leads to an excellent description of the HERA data \cite{Beuf:2020dxl}. On the other hand, T. Stebel discussed prompt photon production within the high energy factorization approach which is suitable for dilute-dilute scattering. He argued that by considering two leading partonic channels, namely $qg^*\to q\gamma$ and $g^*g^*\to q\bar q\gamma$ (with the latter being order of $(\alpha_s)^2$ while former is of order $\alpha_s$) one can finds a sensitivity of the prompt photon transverse momentum distribution to the gluon transverse momentum distribution \cite{Golec-Biernat:2020cah}. Finally,  F. M. Deganutti discussed Mueller-Tang jets at NLO within the high-energy factorization approach. He argued that by including the leading logarithmic BFKL evolution and NLO jet vertex for this specific observable, one gets a discrepancy between the theoretical calculations and the experimental data. He also argued that next-to-leading logarthmic corrections to the BFKL evolution are large and has to be accounted for.

Turning now from NLO corrections to the sub-eikonal corrections, D. Adamiak presented the results of a theory that describes the helicity Parton Distribution Functions (hPDFs) in terms of the polarized dipole amplitudes and shown the first small-x fit to world polarized DIS data. Their predictions for the $g_1$ structure function extends down to $x=10^{-5}$ while maintaining the control over the uncertainty \cite{Adamiak:2021wju}. M. G. Santiago presented his work on the small-x asymptotic of quark Sivers function. He discussed the construction of the transversely polarized Wilson line operators beyond the eikonal accuracy and by expressing the quark Sivers function in terms of these transversely polarized Wilson lines he has shown that main contribution to this quantity is the spin-dependent odderon \cite{Santiago:2021shh}. On the more theoretical side, the talks of G. Beuf and J. Jalilian-Marian were concentrated on inclusive production beyond eikonal accuracy. G. Beuf, presented a systematic derivation of the next-to-eikonal corrections to the quark propagator which are written in a gauge invariant manner, and discussed its effects for forward quark-nucleus scattering \cite{Altinoluk:2020oyd}. J. Jalilian-Marian discussed the quark-photon in proton-nucleus  collisions via helicity amplitudes that involves dynamics of gluon saturation both at small and large $x$ of the target . He also discussed how this formalism can be applied to dijet production in DIS \cite{Jalilian-Marian:2019kaf}.


\subsection{Diffraction}
\label{diffraction}

The theoretical description of diffractive particle production processes that can be studied at an
Electron-Ion Collider (EIC) at Brookhaven National Laboratory in the USA become recently very topical. 
Collisions of electrons and protons as well as electrons and nuclei will be studied in a broad range of center of mass energies from 20 to 140 GeV. Apart from its large luminosity, the EIC will also accelerate light and heavy nuclei, and it will be the first polarized electron-hadron collider ever. 
At DIS2021, we had five talks devoted to the diffractive processes in electron-proton and electron-nucleus 
collisions. A. Stasto discussed the possibilities of measuring inclusive diffraction at EIC
\cite{AbdulKhalek:2021gbh}. Thanks to the excellent forward proton tagging, the EIC will be able to access the wider kinematical range of longitudinal momentum fraction and momentum transfer of the leading proton than at HERA. EIC would allow for precise measurements of longitudinal diffractive structure function and 
for the precise extraction of the nuclear diffractive parton distribution functions \cite{Slominski:2021zit}.

S. Munier presented the results of a theory that describes the cross section for the diffractive dissociation of a small onium off a large nucleus conditioned to a minimum rapidity gap can be identified to a simple classical observable on the stochastic process representing the quantum evolution of the onium in the QCD dipole model. He argued that events in which a large number of dipoles interact simultaneously bring a sizable contribution to the diffractive cross sections, which we are able to characterize quantitatively \cite{Le:2020zpy}. 

Next, we had two talks concentrated on the photoproduction of diffractive dijets. I. Helenius gave a talk on the new framework for the simulations of hard diffractive events in photoproduction within Pythia 8.
He presented the model applies the dynamical rapidity gap survival probability based on the multiparton interaction model in Pythia. He argued that these additional interactions provide a natural explanation for the observed factorization-breaking effects in hard diffraction by filling up the rapidity gaps used to classify the events of diffractive origin \cite{Helenius:2019gbd}. M. Klasen presented the results for diffractive dijet photoproduction at the EIC. He showed that the EIC will provide new and more precise information on the diffractive parton density functions (PDFs) in the pomeron than previously obtained at HERA and provide access to a new nuclear diffractive PDFs \cite{Duwentaster:2021bkb}.

Finally, T. Toll presented his recent studies of intranuclear fluctuations in eA collisions using both the saturated and non-saturated dipole models, which are implemented in the Sartre event generator. He showed 
EIC predictions. These have a sizable effect in the large |t|-region of exclusive diffraction \cite{Sambasivam:2019gdd}.


\subsection{Exclusive processes and Vector Meson production}
\label{VM}

Let us now continue our summary with the talks that were devoted to the exclsuive processes. A. Szczurek presented his recent work about the dilepton production in proton-proton collisions with rapidity gap in the main detector and one forward proton in the forward proton detectors. The calculations are performed including transverse momenta of the virtual photons and using relevant off-shell matrix elements. He showed that the corresponding gap survival factor depends on the invariant mass of the dilepton system as well as the mass of the proton remnant. The gap survival with and without proton measurement in forward proton detector were compared and the underlying dynamics was discussed \cite{Szczurek:2021elt}. Moving on to the exclusive production of vector mesons, F. Celiberto presented results for the exclusive electroproduction of $\rho$ mesons in the high-energy limit of strong interactions, pointing out how ratios of polarized amplitudes are excellent observables to discriminate among several existing UGD models. A high-energy factorization formula is established within BFKL, given as the convolution of an impact factor depicting the forward-meson emission and of an unintegrated gluon distribution (UGD) driving the gluon evolution at small-x. Being a nonperturbative quantity, the UGD is not well known and several models for it have been proposed so far \cite{Bolognino:2018rhb}.

H. Mantysaari discussed his new results for higher-order corrections to exclusive $J/\psi$ production. 
He showed how the first relativistic correction and the NLO correction in the longitudinal polarization case affect exclusive production of $J/\psi$. The relativistic corrections are important for a good description of the HERA data, especially at small values of the photon virtuality. The next-to-leading order results for longitudinal production are evaluated numerically \cite{Mantysaari:2020axf}. Continuing the topic of higher order corrections to exclusive vector meson production, J. Penttala showed new results on relativistic corrections to the heavy vector meson light front wave function and calculation on NLO corrections in $\alpha_S$ to the exclusive vector meson production in the nonrelativistic limit. The NLO corrections are found to have a sizable contribution to the cross section, emphasizing their significance in phenomenological comparisons to the HERA exclusive $J/\psi$ production data \cite{Lappi:2020ufv}.

Next, we had the talk concentrated on exclusive photoproduction of $J/\psi$ mesons in ultraperipheral heavy-ion collisions in the color dipole approach. It is a continuation of previous investigations
\cite{Luszczak:2017dwf,Luszczak:2019vdc}, with a nuclear dipole cross section which is based on
its free-nucleon counterpart obtained through fits to HERA data \cite{Luszczak:2013rxa,Luszczak:2016bxd}.
A. Luszczak discussed the role of $c \bar c g$-Fock states in the diffractive photoproduction
of $J/\psi$-mesons. She showed results based on earlier analysis \cite{Luszczak:2019vdc}, with the incorporation of gluon shadowing corrections. They compared the new calculations to the recent data on the photoproduction of $J/\psi$ by the ALICE and LHCb collaborations and they observed that the inclusion of
$c \bar c g$-Fock states improves agreement of the dipole approach with the midrapidity data of the ALICE collaboration.

M. Krelina studied the momentum transfer dependence of differential cross sections $d\sigma/dt$ 
in diffractive electroproduction of heavy quarkonia on proton targets. He argued that the significance of dipole orientation becomes stronger towards smaller photon energies and can be tested by experiments at the LHC and EIC \cite{Kopeliovich:2021dgx}. I. Babiarz presented new results on central exclusive production of $\eta_c$ and $\chi_c$ in kt- factorization. She revisited the pQCD formulation of central exclusive processes in the example of the production of spinless quarkonia. She proposed a way to calculate the soft effects, in the region of small gluon transverse momenta, using UGDs obtained from color dipole models and a simple prescription for its off-diagonal extrapolation \cite{Babiarz:2020jhy}. Finally, P. Lebiedowicz
presented a new study of the central exclusive diffractive production of axial vector $f_1$ mesons in proton-proton collisions via pomeron-pomeron fusion within the tensor-pomeron approach. He showed how to adjust the parameters of his model to the WA102 experimental data \cite{Lebiedowicz:2020yre}.

\section{Experimental Talks}
\label{experimental_talks}
The DIS WG2 session included presentations from all the major experimental collaborations at RHIC and at the LHC. Results analyzing archived data from H1 and ZEUS were also presented. The experimental program is highly focused around diffractive measurements and studies from ultra-peripheral heavy-ion collisions (UPCs). During the workshop session, it became clear that UPC measurements are providing more precise tests for QCD studies, including low-x physics. In addition, new observables such as exclusive dijets and angular correlations are providing new ways to test models with polarization observables. 

\subsection{Vector meson photoproduction studies}

Tomas Herman presented the latest ALICE results on J/$\psi$ photoproduction in ultra-peripheral collisions at the LHC. The first results of momentum transfer distribution were presented and compared to pQCD models, namely, the nuclear shadowing LTA model and the gluon saturation-based b-BK calculation. Both models are in agreement with the data. William Schmidke presented STAR results for J/$\psi$ production in ultra-peripheral heavy-ion collisions. STARLigth and SARTE were compared to the momentum transfer distribution squared measured by STAR. He explained that a larger data set expected in the coming years will provide more precise comparisons. He also presented performance studies of J/$\psi$ photoproduction with polarized protons. Spencer Klein presented a discussion of possible lessons and issues for future measurements of the momentum transfer distribution in UPCs and at the EIC.

Arthur Bolz presented the measurement of exclusive $\pi^+\pi^-$ and $\rho^{0}$ meson photoproduction with the H1 detector at HERA. This recent study includes measurements for the elastic and proton-dissociative component, and consistent with recent measurements using the CMS detector and old HERA measurements. Detailed $\rho^{0}$ measurements were also performed using the ALICE detector, for coherent photoproduction in ultra-peripheral Pb-Pb and Xe-Xe events. This was presented by Valeri Pozdnyakov. The PbPb results are found to be in agreement with models based on the color-dipole approach and with Gribov-Glauber shadowing calculations. The first Xe-Xe measurements, together with those in PbPb events provide an interesting A-dependence study to probe nuclear shadowing. 

\'Oscar Boente Garc\'ia from the LHCb collaboration presented their first results on J/$\psi$ photoproduction from ultra-peripheral Pb--Pb collisions, as well as the observation of the photoproduction of J/$\psi$ mesons from peripheral Pb--Pb events. 

\subsection{Exclusive dijets with large momentum transfer in UPCs}
Alexander Bylinkin presented the measurement of the angular correlation of exclusive dijet events with large momentum transfer using ultra-peripheral PbPb collision at 5 TeV with the CMS detector. The observed angular decorrelation cannot be explained from purely back-to-back events or expectations from photon-proton models computed from RAPGAP. 

\subsection{Diffraction and other QCD studies in pp collisions}

Using the STAR detector, Leszek Adamczyk presented the measurement of charged-particle production in single diffractive pp events, where the final state protons were reconstructed. Results for both inclusive and identified charged-particle production were presented. Among their results, they reported that their measured antiproton-proton ration suggest contributions from backward baryon transfer are  needed to explain the data.  Also using the STAR detector, Rafal Sikora presented recent results on Central Exclusive Production at 200 GeV and 510 GeV. Several resonance-type structures were observed in the dipon mass spectra, including a resonance around 2.2 GeV. 

Christophe Royon  presented a study comparing pp and $p\bar p$ differential elastic cross section measured by the D0 and TOTEM collaboration, respectively. The data, after some theoretical model considerations, could be explained via the odderon exchange.

Cristian Baldenegro Barrera presented the first measurement of hard color--single exchange in dijet events in 13 TeV pp events with the CMS and TOTEM detectors. The so-called ``jet-gap-jet" events with CMS showed that about 0.5\% of such events are produced by hard color-singlet exchange, and these 13 TeV pp results do not show any further suppression with respect to those previously studied at 7 TeV pp. He presented the first study using the TOTEM detector. Here the hard color-single fraction was found to be larger.

\subsection{Diffraction with nuclei}
Using 8 TeV p-Pb collision data, CMS has studied forward rapidity gap distributions. Dmitry Sosnov presented this measurement that can serve for better tuning the contribution of diffraction in Monte Carlo event generators, particularly useful for cosmic rays physics. The reported results show a significant discrepancy for commonly used MCs. This is also the first time that the inclusive photon-induced process is observed as a function of large rapidity gap at high energies. 

\subsection{Prospects of intact proton measurements}
Fabrizio Ferro from CMS presented proton reconstruction studies with the Precision Proton Spectrometer (PPS), showing performance studies for the LHC Run 2 data, as well as prospects for Run 3. Along this direction, Ksenia Shchelina presented CMS results and prospects for diffraction studies and searchers for physics beyond the standard model using the Precision Proton Spectrometer. She discussed prospects for future di-photon processes at high masses.
The ATLAS collaboration also has a physics program in this direction, and Jesse Liu discussed possible future studies using the ATLAS Forward Proton where processes such the photon fusion process can be studied. 

\subsection{Other studies}
The results from the STAR collaboration on di-hardon correlations in pp, p-Au and p-Al collisions were presented by Xiaoxuan Chu. Such studies are part of the cold QCD program at RHIC energies. At the workshop the first indication that the di-hadron correlations are stronger suppressed in pAu with respect to pAl were presented.
Azimuthal correlations can also be studied in more elementary systems. Amilkar Quintero presented the azimuthal decorrelation between the leading jet and the scattered lepton in DIS with ZEUS data at HERA. Such measurements have been reported for $\gamma$-jet, but never for the lepton-jet channel.

\section{Conclusion}
\label{conclusions}
DIS2021 showed once again that the physics community is very much interested in the physics of small-x and diffractive processes. We witnessed recent progress in the theoretical and experimental programs in many fronts, and is particularly exciting to see the participation of both junior and senior members of the community. Apart from presenting the status of the various programs, new promising ideas and some planned prospects for the coming years were discussed. This year, which was the first time this conference was conducted in an online format, was unique because of the large participation of members of the community submitting abstracts, yet the live in-person discussions, and, the informal interactions after each coffee break or during the social events, were greatly missed. 

\section*{Acknowledgements}

The work of  TA is partially supported by Grant No. 2018/31/D/ST2/00666 (SONATA 14 - National Science Centre, Poland) and by  MSCA RISE 823947 ''Heavy ion collisions: collectivity and precision in saturation physics'' (HIEIC). DTT acknowledges support from the Department of Energy, Office of Science, Nuclear Physics, award number DE-SC0020914

\bibliography{Summary_bib.bib}
\nolinenumbers

\end{document}